\documentstyle[11pt]{article}
\input{epsf}

\begin{document}

\title{\bf Thermal and Electric Conductivities of
           Coulomb Crystals in the Inner Crust of
           a Neutron Star}

\author{ {\bf D.A. Baiko and D.G. Yakovlev} \\
         {\it Ioffe Physical Technical Institute, St.Petersburg} \\
         {\it Received on  November 17, 1995}}

\date{${}$}
\maketitle

\def\la{\;\raise0.3ex\hbox{$<$\kern-0.75em\raise-1.1ex\hbox{$\sim$}}\;}
\def\ga{\;\raise0.3ex\hbox{$>$\kern-0.75em\raise-1.1ex\hbox{$\sim$}}\;}
\newcommand{\om}{\mbox{$\omega$}}              
\newcommand{\th}{\mbox{$\vartheta$}}           
\newcommand{\ph}{\mbox{$\varphi$}}             
\newcommand{\ep}{\mbox{$\varepsilon$}}         
\newcommand{\ka}{\mbox{$\kappa$}}              
\newcommand{\dd}{\mbox{d}}                     
\newcommand{\vp}{\mbox{\boldmath $p$}}         
\newcommand{\vk}{\mbox{\boldmath $k$}}         
\newcommand{\vq}{\mbox{\boldmath $q$}}         
\newcommand{\vv}{\mbox{\boldmath $v$}}         
\newcommand{\vect}[1]{\mbox{\boldmath $#1$}}   
\newcommand{\vF}{\mbox{$v_{\rm F}$}}           
\newcommand{\pF}{\mbox{$p_{\rm F}$}}           
\newcommand{\kF}{\mbox{$k_{\rm F}$}}           
\newcommand{\kTF}{\mbox{$k_{\rm TF}$}}         
\newcommand{\kB}{\mbox{$k_{\rm B}$}}           


\begin{abstract}
Thermal and electric conductivities
of relativistic degenerate electrons are calculated for
the case when electrons scatter by phonons
in Coulomb crystals made of spherical finite--size nuclei
at densities $10^{11}$~g/cm$^3 \la \rho \la 10^{14}$~g/cm$^3$,
corresponding to the inner crust of a neutron star.
In combination with the results of the previous article
(for lower $\rho$), simple unified fits are obtained
which describe the kinetic coefficients in the range
$10^3$~g/cm$^3 \la \rho \la 10^{14}$~g/cm$^3$,
for matter with arbitrary nuclear composition.
The results are valid for studying thermal evolution of neutron
stars and evolution of their magnetic fields.
The difference between the kinetic coefficients
in the neutron star crust composed of ground state
and accreted matters is analyzed. Thermal drift of the magnetic field
in the neutron star crust is discussed.
\end{abstract}

\newpage

\section{INTRODUCTION}
The thermal end electrical conductivities of degenerate electrons
which suffer electron--phonon scattering in Coulomb crystals
of atomic nuclei in dense matter of white dwarfs and neutron stars
has been studied by a number of authors. In particular,
Yakovlev and Urpin (1980) developed an approximate method
of evaluation of the kinetic coefficients which takes into account
the main features of phonon spectrum in the body-centered-cubic
Coulomb crystals and the leading role of the Umklapp processes
in electron--phonon scattering under astrophysical conditions.
The same authors presented critical analysis of the preceding
papers. Raikh and Yakovlev (1982) confirmed the validity of
the approximate method by direct Monte Carlo calculations
of the kinetic coefficients with allowance for the exact phonon
spectrum and contributions from the normal and Umklapp scattering
processes. Itoh et al. (1984, 1993) used the same approximate method
but included the Debye -- Waller factor. These authors fitted their
results by very complicated expressions. Finally, Baiko and Yakovlev
(1995) calculated the kinetic coefficients
directly by the Monte Carlo method and by
the approximate analytic method, including the Debye -- Waller factor,
in the density range $10^3$ g/cm$^3 \la \rho \la 10^{11}$ g/cm$^3$,
which corresponds to the cores of white dwarfs and outer crusts
of neutron stars. Monte Carlo simulations confirmed the validity
of the results of Itoh et al. (1984, 1993) obtained by the
approximate method. The analytic approach allowed Baiko and Yakovlev
(1995) to fit the kinetic coefficients by simple equations,
valid for any nuclear composition, and to extend the
results to the case of the face-centered-cubic Coulomb crystals,
which could also exist in dense stellar matter but which
had not been considered earlier.

The aim of the present article is to study the kinetic coefficients
at higher densities $10^{11}$ g/cm$^3 \la
\rho \la 10^{14}$ g/cm$^3$. This range corresponds to the
inner crust of a neutron star where finite sizes of atomic nuclei
are important. Combining the results of the present article and
the analytic fits of Baiko and Yakovlev (1995) for lower densities,
we will obtain simple fit expressions valid equally for the outer and
inner crusts of neutron stars with any nuclear composition.

\section{GENERAL RELATIONSHIPS}

Consider cold matter in the density range
$10^{11}$ g/cm$^3 \la \rho \la 10^{14}$ g/cm$^3$.
At $\rho < \rho_{\rm d}$ ($\rho_{\rm d} \approx (4 \,-\,6) \times 10^{11}$
g/cm$^3$ is the neutron drip density) matter consists of nuclei
($Z,A$) and strongly degenerate, ultrarelativistic and almost
ideal electrons. We assume that, at
fixed $\rho$ and $T$, there are nuclei of single species.
Highly energetic electrons induce $\beta$
captures and, therefore, large neutron excess of atomic nuclei
(e.g., Shapiro and Teukolsky, 1983). At higher densities,
$\rho_{\rm d} < \rho \la 10^{14}$~g/cm$^3$, matter
contains also free neutrons in addition to the nuclei and electrons.
With the grows of $\rho$, the fraction of the neutrons
increases. At $\rho \ga 10^{14}$ g/cm$^3$ the nuclei
may become nonspherical and/or form clusters
(Lorenz et al., 1993). The nuclear composition
depends on pre-history of the neutron star crust. For instance, the
composition of ground state matter
(Negele and Vautherin, 1973; Haensel and Pichon, 1994)
differs noticeably from that of accreted matter
(Haensel and Zdunik, 1990 a, b).

The properties of matter depend also on temperature $T$.
At low $T$ the nuclei (ions) form Coulomb crystal. One commonly studies
body-centered-cubic crystals which are most tightly bound.
However the difference of binding energies of the body-centered-cubic
and face-centered-cubic crystals is very small, and one cannot exclude
the presence of face-centered-cubic crystals in the neutron star
crust (e.g., Baiko and Yakovlev, 1995). The classical
body-centered-cubic crystal melts when the ion coupling parameter
$\Gamma \equiv Z^2 e^2/(a \kB T)$ reaches the critical value
$\Gamma = \Gamma_{\rm m} \approx 172$ (Nagara et al., 1987).
In this case $a = [3/(4 \pi n_{\rm i})]^{1/3}$ is the mean
inter-ion distance, $n_{\rm i}$ is the number density of the nuclei,
and $\kB$ is the Boltzmann constant. Accordingly, the melting
temperature is (Figures 1 and 2)
\begin{equation}
         T_{\rm m}  =  {Z^2 e^2 \over
               a \kB \Gamma_{\rm m}} \approx
               1.323 \times 10^5 Z^{5/3} \,
               \left( {\rho_6 \over \mu_{\rm e}} \right)^{1/3}
               {172 \over \Gamma_{\rm m}} \; {\rm K},
\label{1}
\end{equation}
where $\mu_{\rm e}$ is the number of baryons per one electron,
$\rho_6$ is density in units of $10^6$ g/cm$^3$.

The thermal and electric conduction in crystalline matter
($T<T_{\rm m}$) is mainly provided by the electrons due to the
electron--phonon scattering. Under astrophysical conditions,
the normal and Umklapp scattering processes are allowed
(e.g, Raikh and Yakovlev, 1982). The scattering regime
is determined by the relationship between
the temperature $T$ and the ion plasma temperature (Figures 1 and 2)
\begin{equation}
       T_{\rm p}  =  {\hbar \omega_p \over \kB} \approx
               7.832 \times 10^6 \,
               \sqrt{{Z \rho_6 \over A \mu_e}} \; {\rm K},
\label{2}
\end{equation}
where $\omega_{\rm p} = \sqrt{4 \pi Z^2 e^2 n_{\rm i}/m_{\rm i}}$
is the ion plasma frequency and $m_{\rm i}$ is the ion mass.
Note that the Debye temperature of the crystal is
$T_{\rm D}=0.45 T_{\rm p}$ (Carr, 1961). If $T \ga T_{\rm p}$,
many thermal phonons are excited, and the electron scattering
can be treated classically (as caused
by thermal lattice vibrations).
At $T \ll T_{\rm p}$ the number of thermal phonons is
strongly reduced, and the scattering is essentially quantum
(involves single phonons). In the both cases the Umklapp processes
dominate the normal ones. However, at very low temperatures
$T \ll T_{\rm U}$ (Figures 1 and 2), the Umklapp processes
are frozen out, and the normal processes become more significant
(e.g., Raikh and Yakovlev, 1982). Here,
$T_{\rm U} \sim T_{\rm p} Z^{1/3} e^2/(\hbar \vF)$, $\vF \approx c$
being the electron Fermi velocity. We shall restrict ourselves
to the case of $T \ga T_{\rm U}$, which is most important for applications.
In this case, one can use the free electron approximation.

We express
the thermal conductivity $\kappa$ and electric conductivity $\sigma$
through the effective electron collision frequencies
$\nu_\kappa$ and $\nu_\sigma$:
\begin{eqnarray}
      \kappa & = & \frac{ \pi^2 \kB^2 T n_{\rm e}}{3 m_\ast \nu_\kappa}
             \approx 4.04 \times 10^{15} x^2 \beta T_6
             \left( \frac{10^{16} {\rm Ó}^{-1}}{\nu_\kappa} \right)
             \; \; {\rm \frac{ergs}{cm \; s \; K}},
      \nonumber \\
      \sigma & = & \frac{ e^2 n_{\rm e}}{m_\ast \nu_\sigma}
             \approx 1.49 \times 10^{22} x^2 \beta
             \left( \frac{10^{16} {\rm Ó}^{-1}}{\nu_\sigma} \right)
             \; \; {\rm \frac{1}{s}}.
      \label{3}
\end{eqnarray}
In this case $x = p_{\rm F}/(m_{\rm e} c)$,
$\beta = \vF/c = x /\sqrt{1+x^2}$,
$m_\ast = m_{\rm e} \sqrt{1+x^2}$, and $p_{\rm F}$ is the electron Fermi
momentum.

The frequencies $\nu_\kappa$ and $\nu_\sigma$
are conveniently expressed
(Yakovlev and Urpin, 1980) through the dimensionless functions
$F_\sigma$ and $F_\kappa$:
\begin{equation}
      \nu_{\sigma,\kappa}  =
          { e^2 \over \hbar \vF} \, {\kB T \over \hbar} F_{\sigma, \kappa}
          \approx 0.955 \times 10^{15}
          {T_6 \over \beta} F_{\sigma, \kappa} \; \; {\rm s}^{-1}.
      \label{4}
\end{equation}
For calculating these functions, we will use the approximate
formulae derived by Baiko and Yakovlev (1995). The general formalism
is independent of lattice type. The cases of the body-centered-cubic
and face-centered-cubic crystals differ through
numerical factors
determined by phonon spectrum. According to Baiko and Yakovlev (1995),
\begin{eqnarray}
    F_\sigma & = & G_0(t) K_0,~~~
    F_\kappa  =  F_\sigma +
       G_2(t)\left(3 K_2 - {1 \over 2} K_0 \right),
\label{5} \\
    K_0 & = &  \int_{q_{\rm min}}^{2k_{\rm F}} \,
       {\hbar^2 q \, \dd q \over p_{\rm F}^2}
       \, { | f(q)|^2 \over |\epsilon (q)|^2}
       \left(1- {\beta^2 \hbar^2
       q^2 \over 4 p_{\rm F}^2} \right) \, {\rm e}^{-2W}
\nonumber \\
            & = & 2 \, \int_{u_0}^1 \,
       \dd u \, { | f(q)|^2 \over |\epsilon (q)|^2}
       \left(1- \beta^2 u \right) \, {\rm e}^{-2W},
\nonumber \\
      K_2 & = & \int_{q_{\rm min}}^{2k_{\rm F}} \,
       {  \dd q \over q}
       \, { | f(q)|^2 \over |\epsilon (q)|^2}
       \left(1- {\beta^2 \hbar^2 q^2 \over 4 p_{\rm F}^2} \right)
       \, {\rm e}^{-2W}
\nonumber \\
       & = & {1 \over 2} \, \int_{u_0}^1 \,
       {\dd u  \over u} \, { | f(q)|^2 \over |\epsilon (q)|^2}
       \left(1- \beta^2 u \right)
       \, {\rm e}^{-2W}.
\label{6}
\end{eqnarray}
Here, $\hbar q$ is the momentum transferred by an electron
due to emission or absorption of a phonon,
$u=[\hbar q /(2 p_{\rm F})]^2$,
$t=T/T_{\rm p}$ is the dimensionless temperature,
${\rm e}^{-2W}$ is the Debye --- Waller factor,
$f(q)$ is the nuclear formfactor,
$\epsilon(q)$ is the static
longitudinal dielectric function of the degenerate
electron gas (Jancovici, 1962). The lower limit of integration over
$q$ ($q=q_{\rm min}$) restricts the integration domain where
the Umklapp processes
are operative (e.g., Raikh and Yakovlev, 1982);
this limit is set equal to the equivalent radius
of the Brillouin zone $ q_{\rm min}  = (6 \pi^2 n_{\rm i})^{1/3}$;
$u_0 = [\hbar q_{\rm min}/(2 p_{\rm F})]^2=1/(4Z)^{2/3}$.

The functions $G_0$ and $G_2$ in Equations
(5) are determined by the phonon spectrum. They have been
calculated and fitted by Yakovlev and Urpin (1980),
Raikh and Yakovlev (1982), and also by Baiko and Yakovlev (1995):
\begin{eqnarray}
   G_0(t) & = & {u_{-2} t \over \sqrt{t^2 + a_0  }},
\nonumber \\
   G_2(t) & = & { t \over \pi^2 (t^2+ a_2)^{3/2} }.
\label{7}
\end{eqnarray}
In this case $u_{-2}$, $a_0$ and $a_2$ are the constants determined by
the lattice type and
$u_{-2}$ is a frequency moment of the phonon spectrum.
For the body-centered-cubic crystal, we have
$u_{-2}^{({\rm bcc})}=13.00$,
$a_0^{({\rm bcc})}=0.0174$ and $a_2^{({\rm bcc})}=0.0118$.
For the face-centered-cubic crystal, $u_{-2}^{({\rm fcc})}=28.80$,
$a_0^{({\rm fcc})}=0.00505$ and $a_2^{({\rm fcc})}=0.00461$.

As shown by Baiko and Yakovlev (1995),
the Debye -- Waller parameter can be fitted
(with the mean error of $\sim$ 1\%) as
\begin{eqnarray}
        2 W(q) & = &
         \alpha u ,
\label{8} \\
        \alpha & = &  \alpha_0  \;
             \left( {1 \over 2} u_{-1} {\rm e}^{-9.100t}+tu_{-2} \right),
\nonumber \\
        \alpha_0 & = & {4 m_{\rm e}^2c^2 \over k_{\rm B} T_{\rm p} m_{\rm i}} 
x^2
                      \approx 1.683 \sqrt{ { x \over AZ} },
\nonumber
\end{eqnarray}
where $u_{-1}$ is another frequency moment of the phonon spectrum,
$u_{-1}^{({\rm bcc})}=2.800$,  $u_{-1}^{({\rm fcc})}=4.03$.

Since we consider the ultrarelativistic electron gas,
it is sufficient to set $x \gg 1$, $\beta = 1$.

For the densities of interest, the sizes of atomic nuclei become
comparable to the inter-ion distance $a$ (see above).
Contrary to the case of $\rho \la 10^{11}$ g/cm$^3$, considered by
Baiko and Yakovlev (1995), the nuclear formfactor $f(q)$
becomes important in the integrals (6). We will not analyze
the nonspherical nuclei which may appear at
$\rho \ga 10^{14}$ g/cm$^3$ (Lorenz et al., 1993),
but restrict ourselves to a standard model of spherical nuclei
with the uniform proton core of radius $r_{\rm c}$:
\begin{eqnarray}
       f(q) & = & \frac{3}{(q r_{\rm c})^3}
               \left[ \sin(q r_{\rm c})
               - q r_{\rm c} \cos(q r_{\rm c}) \right],
\label{9} \\
       q r_{\rm c} & = &  \sqrt{u}  \, \, { r_{\rm c} \over a}
       \left( 18 \pi  Z  \right)^{1/3}.
\nonumber
\end{eqnarray}

\section{NUMERICAL RESULTS AND THEIR APPROXIMATION}

Calculation of the thermal and electric conductivities
reduces to evaluating the integrals $K_0$ and $K_2$
from Equations (6). For the conditions of study,
the functions $K_0$ and $K_2$ depend on three parameters: on the quantity
$\alpha$ in the Debye --- Waller factor (8),
on the nuclear charge number $Z$ and on the parameter
$g= r_{\rm c}/a$ in the nuclear formfactor (9).
We have calculated the integrals (6) at
$Z = 20, 40, 60$, $\alpha$ = 0.04, 0.12, 0.4, 1.2, 4, 12, and
$g$ = 0, 0.1, 0.2, 0.3, 0.4.
Analyzing the properties of matter at $10^{11}$
g/cm$^3 \la \rho \la 10^{14}$ g/cm$^3$,
one can see (Section 4) that the adopted grid
covers all possible parameter domain.

Let us fit the newly calculated
values of $K_\sigma$ and $K_\kappa$ basing on
the analytic expressions for
$F_\sigma$ and $F_\kappa$ at $\rho \la 10^{11}$ g/cm$^3$
obtained by Baiko and Yakovlev (1995). First of all, note that
the electric screening is rather insignificant.
As shown by Baiko and Yakovlev (1995), one can set $\epsilon(q)$=1
in the integrals (6) and include the weak screening effect
by shifting the integration limit
$u_0$ to $u_1=(4Z)^{-2/3}+u_e$, where
$u_e = [\hbar k_{\rm TF} /(2 p_{\rm F})]^2 = e^2/(\pi \hbar \vF)
\approx 1/(137 \pi \beta)$ and $k_{\rm TF}$ is the inverse
length of screening of a charge by the degenerate plasma electrons.

First consider the small--size  nuclei ($g \ll 1$).
Expanding the formfactor up to the terms $\propto g^2$, we obtain
$|f(q)|^2 \approx 1 - 0.2(q r_{\rm c})^2$. Then the integrals
(6) are taken:
\begin{equation}
    K_0  =  2 \Phi_1 - {2 \over 5} \left( 2 p_{\rm F} r_{\rm c} \over \hbar
              \right)^2 \Phi_2,~~~
    K_2  =  {1 \over 2} \Phi_0 - {1 \over 10}
              \left( 2 p_{\rm F} r_{\rm c} \over \hbar
              \right)^2 \Phi_1.
\label{10}
\end{equation}
Here, we have introduced the functions
\begin{eqnarray}
    \Phi_k & = & S_{k-1} - \beta^2 S_k,~~~
    S_k = \int_{u_1}^1 \dd u \, u^k \, {\rm e}^{- \alpha u},
\label{11} \\
   S_{-1} & = & E(w)-E(\alpha),~~~
   S_0={1 \over \alpha} \left( {\rm e}^{-w} - {\rm e}^{- \alpha}
       \right),
\nonumber \\
   S_1 & = & { 1 \over \alpha^2}
       \left[ {\rm e}^{-w} (w+1) - {\rm e}^{-\alpha}(\alpha+1) \right],
\nonumber \\
   S_2 & = & { 1 \over \alpha^3}
       \left[ {\rm e}^{-w} (w^2+2w+2) -
       {\rm e}^{-\alpha}(\alpha^2 + 2 \alpha + 2) \right],
\end{eqnarray}
where $w = \alpha u_1$, $E(x)$ is the integral exponent,
which can be easily calculated
using, for instance, Equation (21) of the article of
Baiko and Yakovlev (1995).

The natural generalization of Equation (10) to the case
of large--size nuclei can be provided by the following expressions:
\begin{eqnarray}
    K_0 & = & 2 \Phi_1 \left[1 +
                  { (18 \pi Z)^{2/3} g^2 \Phi_2
                  \over 5 \Phi_1 P_0 }
                  \right]^{-P_0},
\label{13} \\
    K_2 & = &  {1 \over 2} \Phi_0 \left[1 +
                  { (18 \pi Z)^{2/3} g^2 \Phi_1
                  \over 5 \Phi_0 P_2 }
                  \right]^{-P_2},
\label{14}
\end{eqnarray}
where $P_0$ and $P_2$ are the fit parameters. These expressions
describe quite accurately our numerical results with
$P_0 =  4.787 - 0.0346 \, Z$ and
$P_2 =  2.729 - 0.0204 \, Z$. The mean fit error of
Equation (13) is 2\%; the maximum error 4.7\%
takes place at $\alpha=0.04$, $Z=60$ and $g=0.3$.
The mean fit error of (14) is
2.3\%, and the maximum error 10\% takes place at $\alpha=12$,
$Z=60$ and $g=0.4$.

If $g \ll 1$, Equations (13) and (14)
reproduce the fit expressions
(18) -- (27) of the article of Baiko and Yakovlev (1995).
The latter equations are valid at rather low densities
$10^3$ g/cm$^3 \la  \rho \la 10^{11}$ g/cm$^3$,
at which the finite sizes of the nuclei are unimportant.
Thus Equations (13) and (14) cover very large
density range $10^3$ g/cm$^3 \la \rho \la 10^{14}$ g/cm$^3$
(outer and inner neutron star crusts) and are valid for
any nuclear composition.

\section{ELECTRIC AND THERMAL CONDUCTIVITIES}

Let us analyze the electric and thermal conductivities of
electrons in crystalline matter of a neutron star crust.
The nuclear composition of matter is not definitely known.
For illustration, consider two well established models:
ground state matter and accreted matter.

Ground state matter corresponds to the minimum of free energy
per one baryon. This matter can be produced
from hot dense matter of a newly born neutron star in the course of
the subsequent stellar cooling. At densities
$\rho < \rho_{\rm d} \approx 4 \times 10^{11}$ g/cm$^3$
(below the neutron drip, Section 2) we will use
the composition of the ground state matter
calculated by Haensel and Pichon (1994) on the basis of new laboratory
measurements of masses of nuclei with large neutron excess.
At $\rho > \rho_{\rm d}$ we will use the results
of Negele and Vautherin (1973) derived
with the aid of a modified Hartree--Fock method.

The nuclear composition of accreted matter is determined
(Haensel and Zdunik, 1990a, b) by
nuclear transmutations (pycnonuclear reactions,
$\beta$-captures, emission and absorption of neutrons, etc)
in cold dense accreted matter sinking
within the star. The results are practically insensitive
to the accretion rate and initial chemical composition.
The accreted matter is assumed to burn into
$^{56}$Fe at $\rho \la 10^7$ g/cm$^3$.

Both models of matter are formally calculated at $T=0$, but they
are actually valid at $T \la 4 \times 10^9$~K
(as long as the thermal effects do not influence the properties of nuclei).
It is adopted that single species of nuclei is present at every
pressure (density). This leads to jumps of the nuclear composition
$(A,Z)$ with the growth of pressure (density) and to jumps
of typical temperatures depicted in Figures 1 and 2.
At $\rho \la 10^7$~g/cm$^3$ matter consists of
$^{56}$Fe in both models. At higher densities, the compositions
of ground state and accreted matters are noticeably different.
The accreted matter consists of lighter nuclei with lower
$Z$. For instance, at $\rho = 4 \times 10^{12}$ g/cm$^3$
the accreted matter contains neutron rich magnesium nuclei,
$A=44$, $Z=12$, while the ground state matter contains tin nuclei,
$A=159$, $Z=50$. The difference of compositions affects the thermodynamic
and kinetic properties of matter. For example, at
$10^{11}$ g/cm$^3 \la \rho \la 10^{13}$ g/cm$^3$ the melting
temperature of the accreted matter is a factor of
3 --- 10 lower, than that of the ground state matter.

One should know the proton core radius
$r_{\rm c}$, for calculating the thermal and electric conductivities.
Following Itoh et al. (1984) we set:
$ r_{\rm c} = 1.15 \, A^{1/3}$ fm at $ \rho < \rho_{\rm d}$;
$ r_{\rm c} = 1.83 \, Z^{1/3}$ fm at $ \rho > \rho_{\rm d}$.
It is easy to verify that
$g= r_{\rm c} /a \la 0.2$ for both models.
Figure 3 shows the functions $F_\sigma$ and $F_\kappa$
versus $T/T_{\rm p}$ for a body-centered-cubic lattice of
nuclei $A=159$, $Z=50$ ($\rho = 4 \times 10^{12}$ g/cm$^3$).
The curves are qualitatively the same as in matter of lower density
($\rho < 10^{11}$ g/cm$^3$; see Baiko and Yakovlev, 1995).
The allowance for the nuclear size decreases $F_{\sigma, \kappa}$ and
the effective electron collision frequencies $\nu_{\sigma, \kappa}$,
i.e., increases the electric and thermal conductivities.
However, the effect is not strong, about (20 --- 30)\%. This is
clearly seen from the integrals (6):
the nuclear formfactor reduces the
integrands with the growth of $g$
only at large dimensionless momentum transfers
$u \approx 1$ (i.e., at large--angle scattering),
at which the integrands themselves are sufficiently small.
The latter smallness is due to the Debye--Waller factor
(which suppresses collisions with large momentum transfers)
and due to the factor $(1- \beta^2 u)$, that describes
reduction of backward scattering of relativistic electrons.

Figure 4 shows the dependence of $F_{\sigma, \kappa}$
on $T/T_{\rm p}$ for the same conditions as in Figure
3, but for the face-centered-cubic lattice.
As in a lower--density matter (Baiko and Yakovlev, 1995),
the lattice type affects strongly the kinetic coefficients.
The finite sizes of the nuclei again influence the result
weakly, as in the body-centered-cubic lattice.

Figures 5 and 6 display the density dependence of the electric and thermal
conductivities for the ground state and accreted matters
at $T= 10^8$ and $5 \times 10^8$~K. At low $\rho$ the curves are
broken in melting points (Figures 1 and 2). The discontinuities of
the kinetic coefficients are associated with the jumps of
nuclear compositions (see above). At
$10^{12}$ g/cm$^3 \la \rho \la 10^{13}$ g/cm$^3$
the electric and thermal conductivities of the accreted matter are
a factor of 2 --- 3 higher, than those of the ground state matter.
This is naturally explained by lower nucleus charge
(and lower electron collision frequencies) in the accreted matter.

Note that earlier the thermal and electric conductivities
of the ground state matter in the inner crust of a neutron star
was calculated and fitted by Itoh et al. (1984).
Our results agree mainly with the results of these authors.
However, there is a 30 --- 50 \% difference in the thermal
conductivity at densities $\rho \ga 10^{13}$ g/cm$^3$.
Let us emphasize that our fit expressions are much simpler
than those of Itoh et al. (1984), and they are valid
for matter with any nuclear composition.

\section{THERMAL DRIFT OF THE MAGNETIC FIELD}
If the magnetic field is present in the stellar plasma,
various thermal magnetic effects may operate
(Urpin and Yakovlev, 1980). For instance, the Hall component
of the thermopower affects the thermal flux, that emerges from
the star, and induces the thermal drift of the magnetic field.
If the field is not too large,
so that the electron gas is weakly magnetized
($\omega_{\rm B} ~ \tau  \ll 1$, where $\omega_{\rm B} = eB/(cm_\ast)$),
one can neglect the back reaction of the field onto the thermal flux.
According to Equations (5) and (7), at $T \ga T_{\rm p}$
one has $F_\sigma \approx F_\kappa$. This allows us to introduce
the electron relaxation time
$\tau_0(\varepsilon)=\nu_{\sigma}^{-1}=\nu_{\kappa}^{-1}$,
which is the same for the thermal and electric conductions
($\varepsilon$ being the electron energy). Under formulated conditions,
the drift velocity $u$ is parallel to the thermal flux $Q$
(Urpin and Yakovlev, 1980)
\begin{eqnarray}
    u & = & {2 \eta Q \over n_{\rm e} p_{\rm F} v_{\rm F}}
        \approx 72 \eta {\sqrt{1+x^2} \over x^5}
         \left( {T_e \over 10^6 \;{\rm K} }  \right)^4 \;\;
        {\rm m \over yr} ,
\label{15} \\
    \eta & = & {1 \over 2} \; \left. {
            \partial \ln(\tau_0(\varepsilon) \varepsilon^{-1})
           \over \partial \ln p} \right|_{p=p_{\rm F}}.
\label{16}
\end{eqnarray}
Here, we have assumed that
$Q=\sigma_0 T_e^4$, where $T_e$ is the effective temperature
of the stellar surface and $\sigma_0$ is the Stefan -- Boltzmann constant.
The factor $\eta
\sim 1$ and can change sign, that corresponds to
different thermal drift directions. Taking the derivative, we have
\begin{eqnarray}
   \eta & = & 1.5-\beta^2 - {2 G_0(t) \over F_\sigma } \left[ (1-\beta^2)
        e^{-\alpha} { \left| f(2 p_{\rm F}/\hbar) \right|^2
                \over \left| \epsilon(2 p_{\rm F}/\hbar) \right|^2} +
                \beta^4 \int_{u_0}^1 \,
       \dd u \, { \left| f(q) \right|^2
                \over \left| \epsilon(q) \right|^2} \,
                u \, {\rm e}^{- \alpha u}\right].
\label{17}
\end{eqnarray}
The dashed line in Figures 1 and 2 shows those densities and
temperatures at which $\eta$ = 0. One has
$\eta > 0$ above this line, i.e., the thermal drift
is directed outward, and
$\eta < 0$ below the line, so that the magnetic field drifts inward.

Note that Equations (15) and (16)
are actually valid only at $T_{\rm p} \la T \la T_{\rm m}$,
and the dashed line is broken when $T$ reaches $T_{\rm m}$
with the growth of $\rho$.
It is seen that the thermal drift can be directed either
inward or outward in the above temperature range
and at densities up to $10^{10}$ -- $10^{11}$ g/cm$^3$.
At higher densities and $T \ga T_{\rm p}$, the factor
$\eta$ remains positive for the ground state matter.
As for the accreted matter, $T_{\rm m}$ exceeds
$T_{\rm p}$ at these densities, and Equation (16) becomes invalid.
Calculation of the thermal drift at
$T \la T_{\rm p}$ is outside the scope of the present article.

\section{CONCLUSIONS}
We have calculated the thermal and electric conductivities
of relativistic degenerate electrons due to
electron--phonon scattering in Coulomb crystals
of atomic nuclei at densities $10^{11}$ g/cm$^3 \la \rho \la
10^{14}$ g/cm$^{3}$, corresponding to the inner crust of a
neutron star. Calculations are done with the aid of the
approximate analytic method (Baiko and Yakovlev, 1995)
taking exact spectrum of phonons and the
Debye--Waller factor into account. In addition, we have
taken into consideration finite sizes of atomic nuclei.

The results are fitted by a few simple equations which
reproduce also the fit expressions obtained by
Baiko and Yakovlev (1995) for lower densities.
Thus we have obtained the unified fits
which describe the thermal and electric conductivities
in the wide density range $10^3$ g/cm$^3 \la \rho \la 10^{14}$ g/cm$^3$
(in the inner and outer crusts of neutron stars)
below the melting temperature but above
the temperature where the Umklapp processes are frozen out,
for the body-centered-cubic and face-centered-cubic crystals.
Note that the same equations are valid in the presence
of a weak magnetic field ($\omega_{\rm B} ~ \tau  \ll 1$).

Our results can be useful for numerical simulations of thermal evolution
of neutron stars (cooling, nuclear burning of accreted matter)
and evolution of their magnetic fields (ohmic dissipation,
generation due to thermal magnetic effects).

\begin{center}
      {\bf ACKNOWLEDGEMENTS}
\end{center}
This work was partly supported by the
Russian Foundation for Basic Researches (grant No. 93-02-2916),
the International Science Foundation (grant No. R6-A000), and INTAS
(grant No. 94-3834).
One of the authors (D.A. Baiko) is also grateful
to the International Science Foundation the for
undergraduate grant No. 555s.


\newpage

\begin{center}
                   {\bf REFERENCES}
\end{center}

\noindent
Baiko, D.A. and Yakovlev, D.G., Astron. Lett., 1995, vol. 21, p. 702.

\noindent
Carr, W.J., Phys. Rev., 1961, vol. 122, p. 1437.

\noindent
Haensel, P. and Zdunik, J.L., Astron. Astrophys.,
     1990a, vol. 227, p. 431.

\noindent
Haensel, P. and Zdunik, J.L., Astron. Astrophys.,
     1990b, vol. 229, p. 117.

\noindent
Haensel, P. and Pichon, D., Astron. Astrophys.,
     1994, vol. 283, p. 313.

\noindent
Itoh, N., Kohyama, Y., Matsumoto, N., and Seki M.,
      Astrophys. J., 1984, vol. 285, p. 758; erratum vol. 404, p. 418.

\noindent
Itoh, N., Hayashi, H., Kohyama, Y.,
      Astrophys. J., 1993, vol. 418, p. 405.

\noindent
Jancovici, B., Nuovo Cimento, 1962, vol. 25, p. 428.

\noindent
Lorenz, C.P., Ravenhall, D.G., and Pethick C.J.,
     Phys. Rev. Lett., 1993, vol. 70, p. 379.

\noindent
Nagara, H., Nagata, Y., and Nakamura, T.,
     Phys. Rev., 1987, vol. A36, p. 1859.

\noindent
Negele, J.W. and Vautherin, D., Nucl. Phys., 1973,
     vol. A207, p. 298.

\noindent
Raikh M.E. and Yakovlev D.G., Astrophys.
     Space Sci., 1982, vol. 87, p. 193.

\noindent
Shapiro, S.L. and Teukolsky, S.A., Black Holes, White
  Dwarfs, and Neutron Stars, New York: Wiley-Interscience, 1993.

\noindent
Urpin, V.A. and Yakovlev, D.G., Sov. Astron., 1980, vol. 24, p. 325.

\noindent
Yakovlev, D.G. and Urpin, V.A., Sov. Astron., 1980, vol. 24, p. 303.

\newpage

\begin{figure}[t]
\epsfxsize=0.8\hsize
\centerline{{\epsfbox{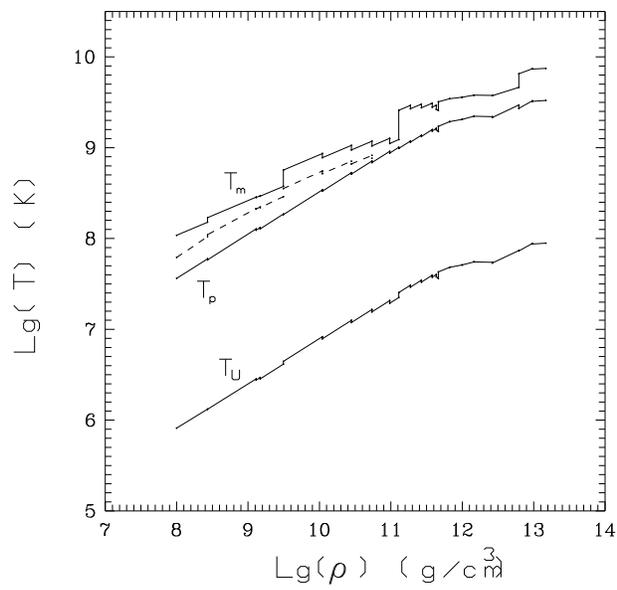}}}
\caption{\label{figure1}
Density --- temperature diagram for the ground state matter:
$T_{\rm m}$ is the melting temperature,
$T_{\rm p}$ is the ion plasma temperature,
$T_{\rm U}$ is the freezing temperature of Umklapp processes.
Dashed line corresponds to the conditions at which the
thermal drift velocity equals zero.
}
\end{figure}

\newpage

\begin{figure}[t]
\epsfxsize=0.8\hsize
\centerline{{\epsfbox{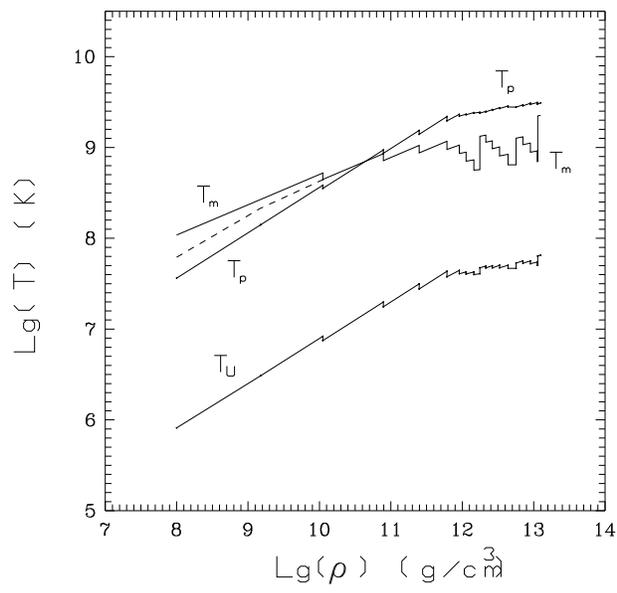}}}
\caption{\label{figure2}
Same as in Figure 1 but for the accreted matter.
}
\end{figure}

\newpage

\begin{figure}[t]
\epsfxsize=0.8\hsize
\centerline{{\epsfbox{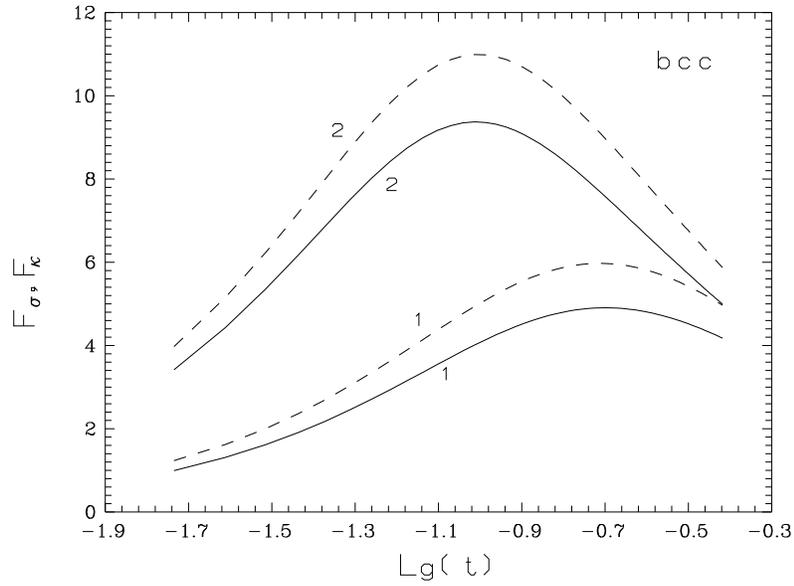}}}
\caption{\label{figure3}
Factors $F_\sigma$ (curves 1) and $F_\kappa$
(curves 2) versus $t=T/T_{\rm p}$
for the body-centered-cubic and face-centered-cubic crystals composed of
$^{159}$Sn nuclei at density
$\rho= 4 \times 10^{12}$ g/cm$^3$ with account for the
finite nucleus size ($g=0.15$, solid lines) and for the point--like nuclei
($g=0$, dashes).
}
\end{figure}

\newpage

\begin{figure}[t]
\epsfxsize=0.8\hsize
\centerline{{\epsfbox{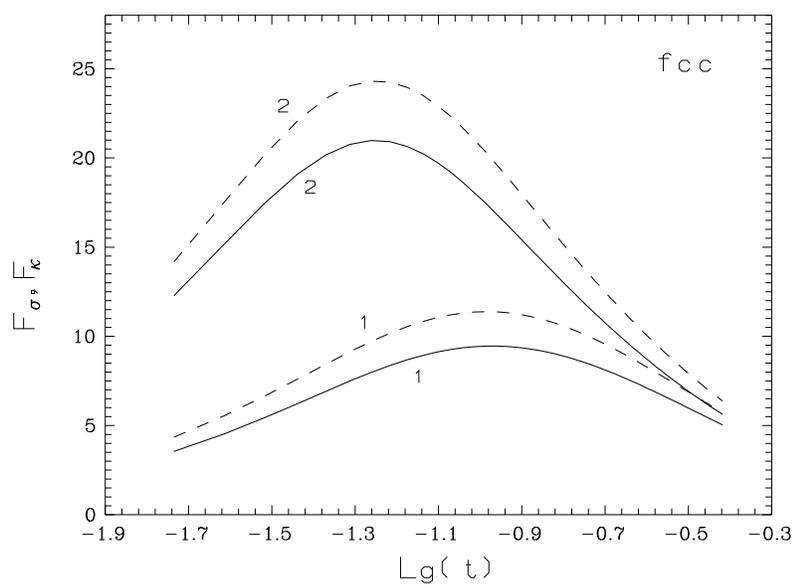}}}
\caption{\label{figure4}
Same as in Figure 3 but for the face-centered-cubic crystal.
}
\end{figure}

\newpage

\begin{figure}[t]
\epsfxsize=0.8\hsize
\centerline{{\epsfbox{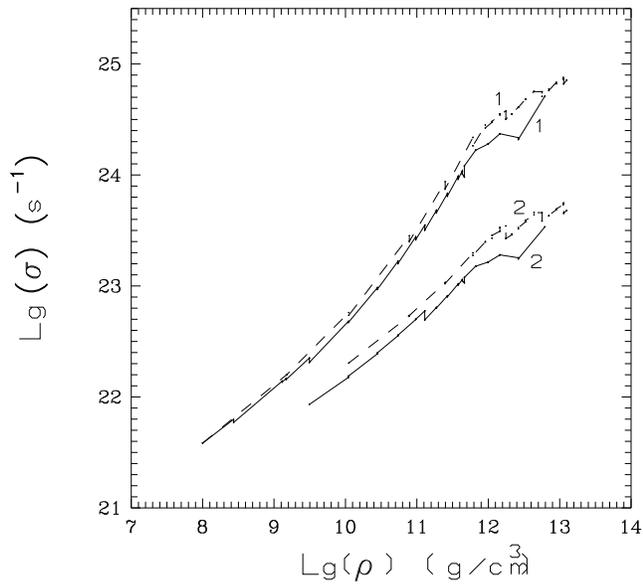}}}
\caption{\label{figure5}
Electric conductivity versus density for the
ground state (solid lines) and accreted (dashes) matters at
$T=10^8$~K (curves 1) and $T=5 \times 10^8$~K (curves 2).
}
\end{figure}

\begin{figure}[t]
\epsfxsize=0.8\hsize
\centerline{{\epsfbox{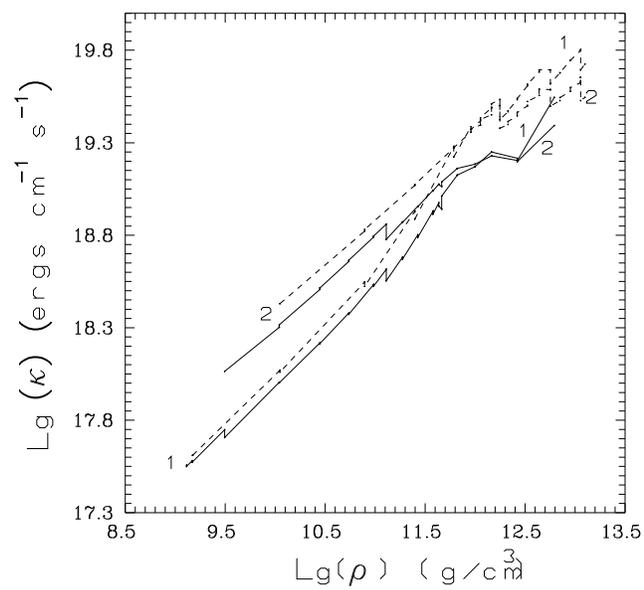}}}
\caption{\label{figure6}
Thermal conductivity versus density for the same conditions
as in Figure 5.
}
\end{figure}

\end{document}